\documentclass[tradiabstract]{aa}
\usepackage{graphicx}
\usepackage{amsbsy}
\usepackage{amsmath}
\usepackage{natbib}
\usepackage{color}
\usepackage{txfonts}
\usepackage{url}
\usepackage{bm}

\DeclareMathSymbol{\varOmega}{\mathord}{letters}{"0A}
\DeclareMathSymbol{\varSigma}{\mathord}{letters}{"06}
\DeclareMathSymbol{\varPsi}{\mathord}{letters}{"09}
\DeclareMathSymbol{\varPhi}{\mathord}{letters}{"08}
\DeclareMathSymbol{\varGamma}{\mathord}{letters}{"00}

\begin{document}

\title{Nucleation and growth of iron pebbles\\ explains the formation of
iron-rich planets akin to Mercury}
\titlerunning{Nucleation of iron pebbles in protoplanetary discs}

\author{Anders Johansen\inst{1,2} \& Caroline Dorn\inst{3}}
\authorrunning{Johansen \& Dorn}

\institute{$^1$ Center for Star and Planet Formation, GLOBE Institute,
University of Copenhagen, \O ster Voldgade 5-7, 1350 Copenhagen, Denmark \\ $^2$ Lund Observatory, Department of Astronomy and Theoretical
Physics, Lund University, Box 43, 221 00 Lund, Sweden \\
$^3$ Institute of Computational Sciences, University of
Zurich, Winterthurerstrasse 109, 8057, Zurich, Switzerland \\e-mail:
\url{Anders.Johansen@sund.ku.dk}}

\date{}

\abstract{The pathway to forming the iron-rich planet Mercury remains
mysterious.  Mercury's core makes up 70\% of the planetary mass, which implies a
significant enrichment of iron relative to silicates, while its mantle is
strongly depleted in oxidized iron. The high core mass fraction is traditionally
ascribed to evaporative loss of silicates, e.g.\ following a giant impact, but
the high abundance of moderately volatile elements in the mantle of Mercury is
inconsistent with reaching temperatures much above 1,000 K during its formation.
Here we explore the nucleation of solid particles from a gas of solar
composition that cools down in the hot inner regions of the protoplanetary disc.
The high surface tension of iron causes iron particles to nucleate homogeneously
(i.e., not on a more refractory substrate) under very high supersaturation. The
low nucleation rates lead to depositional growth of large iron pebbles on a
sparse population of nucleated iron nano-particles. Silicates in the form of
iron-free MgSiO$_3$ nucleate at similar temperatures but obtain smaller sizes
due to the much higher number of nucleated particles. This results in a chemical
separation of large iron particles from silicate particles with ten times lower
Stokes numbers. We propose that such conditions lead to the formation of
iron-rich planetesimals by the streaming instability. In this view, Mercury
formed by accretion of iron-rich planetesimals with a sub-solar abundance of
highly reduced silicate material.  Our results imply that the iron-rich planets
known to orbit the Sun and other stars are not required to have experienced
mantle-stripping impacts.  Instead their formation could be a direct consequence
of temperature fluctuations in protoplanetary discs and chemical separation of
distinct crystal species through the ensuing nucleation process.}


\maketitle

\section{Introduction}

The formation of Mercury from iron-rich planetesimals that were separated
physically from their iron-poor counterparts was discussed in
\cite{Weidenschilling1978} who proposed that the higher density of iron
planetesimals leads to lower orbital decay speeds. However, such aerodynamical
separation does not address how iron and silicates became separated into
distinct planetesimal populations in the first place. A possible pathway to
achieve formation of iron-enhanced planetesimals is that pebbles become
increasingly iron-rich as they grow, due to enhanced growth of dust aggregates
that have a large fraction of iron compared to silicates
\citep{Hubbard2014,KrussWurm2018,KrussWurm2020}.

The alternative evaporation model requires Mercury to have become thermally
processed at ambient protoplanetary disc temperatures above 3,000 K after the
formation of the planetary core \citep{Cameron1985}. One or more giant impacts
between Mercury and other planetary bodies could have caused similar
temperatures and evaporation rates after the dissipation of the protoplanetary
disc \citep{Benz+etal2007,Chau+etal2018}, but this model is challenged by the
subsequent reaccretion of the evaporated silicates \citep{Benz+etal1988},
although the solar wind could reduce the efficiency of the reaccretion
\citep{SpaldingAdams2020}.  Gamma-ray spectroscopy by the MESSENGER probe
revealed a K/Th ratio in Mercury's mantle that is indistinguishable from that of
Venus, Earth and Mars \citep{Peplowski+etal2011}. Since K is a moderately
volatile element with a condensation temperature of around 1,000 K, Mercury
could not have experienced temperatures much in excess of this threshold during
its formation. This finding is thus in tension with the evaporation model.
\begin{figure*}
  \begin{center}
    \includegraphics[width=0.7\linewidth]{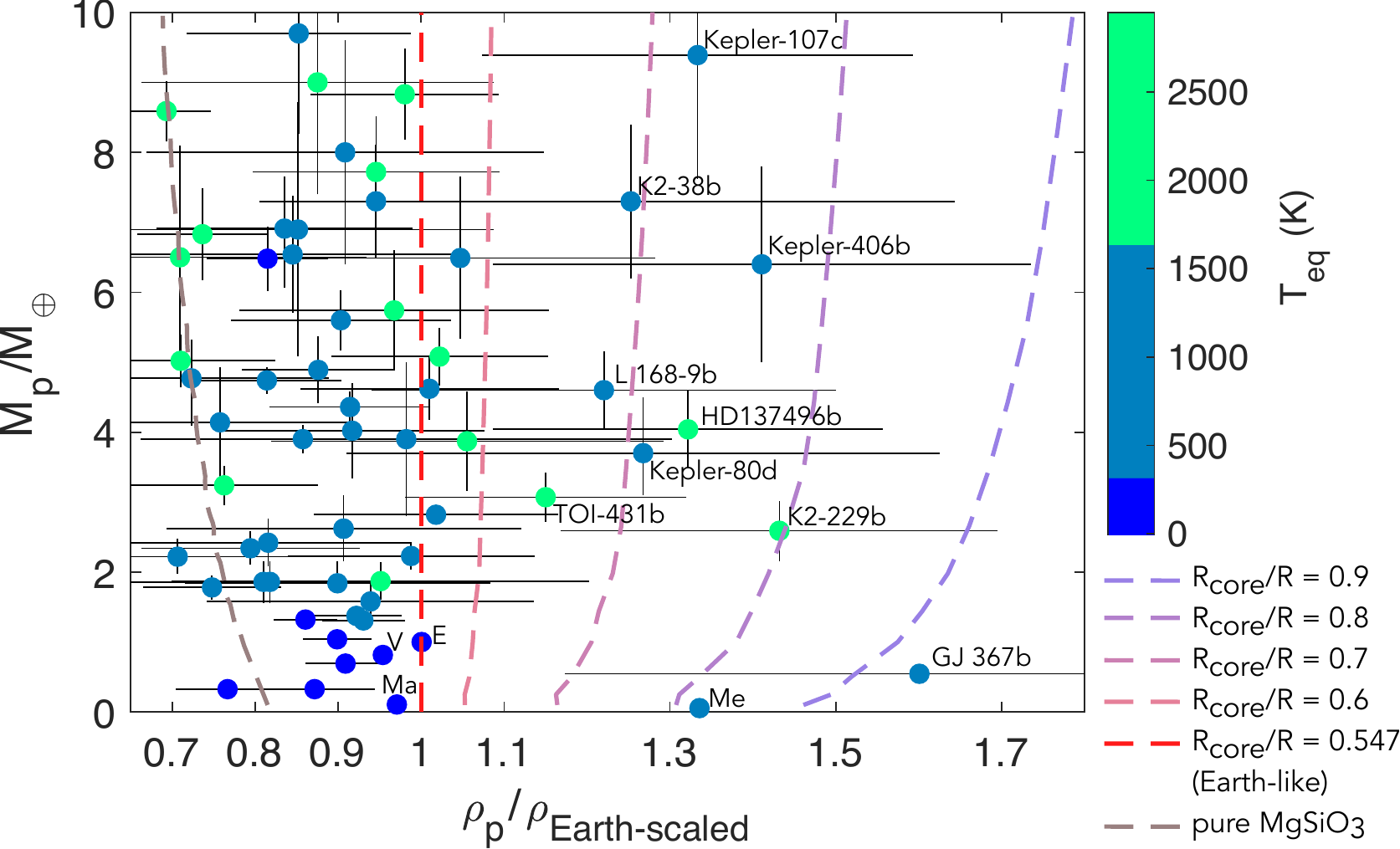}
  \end{center}
  \caption{Scaled densities and masses of observed exoplanets in comparison to
  curves of constant core radius fractions. Planet densities are scaled relative
  to an Earth-analog of fixed core radius fraction (0.547) and for a given mass.
  The selected exoplanets are from the PlanetS database
  \citep{Otegi+etal2020} plus K2-38b and Kepler-406 b (see main text). Earth
  (E), Venus (V), Mars (Ma), and Mercury (Me) are also shown. The emerging
  population of super-Mercuries with densities indicative of iron enrichment
  covers a range of masses up to 5--10 $M_{\rm E}$.}
  \label{f:MRlabel}
\end{figure*}

The highly reduced mantle of Mercury is arguably as puzzling as its high core
mass fraction. The measured fraction of oxidized FeO in the mantle is consistent
with $0\%$ \citep{WarellBlewett2004,VanderKaadenMcCubbin2015,EbelStewart2018},
while Venus, Earth, Mars and Vesta have significant FeO fractions (8\%, 8\%,
18\% and 20\%, respectively) in their crusts and mantles
\citep{RobinsonTaylor2001}.  Silicate minerals with low FeO contents are
expected to condense out under the high pressures prevalent in the inner regions
of the solar protoplanetary disc \citep{KozasaHasegawa1988}. However,
maintaining a low-FeO mantle requires additionally that the contribution to
Mercury of planetesimals formed in the Venus-Earth-Mars region was negligible.
Hence, there must have been very little mixing between the planetesimal
populations that formed the terrestrial planets.  This requires limited stirring
by Jupiter, in strong contrast with models where Jupiter penetrates the
terrestrial planet region during its early migration \citep{Walsh+etal2011}.
This scenario where Jupiter does not stir the planetesimals in the terrestrial
planet zone nevertheless makes it challenging to reach Mars' current mass in
$N$-body simulations \citep{Woo+etal2022}. In the alternative pebble accretion
model for terrestrial planet formation \citep{Johansen+etal2021}, the drifting
pebbles contain largely reduced iron that is later oxidized by dissolution in
hot water within the growing Venus, Earth and Mars \citep{Johansen+etal2022}.
These reduced pebbles would, if they contributed significant mass to Mercury,
therefore agree with the lack of FeO in the crust of Mercury but their solar
Fe/Si value would be inconsistent with Mercury's bulk composition.

The discovery of exoplanets with similar densities as Mercury makes the problem
of Mercury formation even more acute \citep{Santerne+etal2018}.
Fig.~\ref{f:MRlabel} shows observed super-Mercuries with well-characterized
masses and radii; they have masses ranging between 0.5 and 10 $M_{\rm E}$. The
displayed planets are
well-characterized in the sense that they are included in the PlanetS database
\citep{Otegi+etal2020}; in addition we show K2-38b and Kepler-406 b, which are
discussed in \cite{Adibekyan+etal2021b}. In an attempt to explain the high
densities and high iron contents of super-Mercuries several mechanisms were
suggested \citep{Wurm+etal2013, KrussWurm2018, Rappaport+etal2012}, including
collisions \citep{StewartLeinhardt2012,ClementChambers2021}. In contrast to
previous suggestions from impact studies \citep{Marcus+etal2010}, observed
super-Mercuries are not limited to masses of 5 $M_{\rm E}$. Recent formation
studies that incorporate high energy collisions are unable to reproduce the
highest density super-Mercuries \citep{Scora+etal2020}. Further, observations
indicate that super-Mercuries may be a distinct population from super-Earths
with regard to their inferred Fe/(Mg+Si) contents \citep{Adibekyan+etal2021a}.
The scarcity of planets between both populations can suggest that giant impacts
are unlikely the only cause of forming super-Mercuries. This is because the
stochastic nature of collisions should result in a continuous population.
However, this scarcity is less clear in the density-mass space and planetary
data is too limited to draw a conclusion from this speculation.  Overall, it
seems plausible to believe that additional mechanisms other than collisions are
at play to form iron-enhanced worlds. These mechanisms should be able to form
planets at the observed large range of masses and core-radius fractions
(Fig.~\ref{f:MRlabel}).

In this paper we explore how metallic iron separates from iron-free silicates
following cooling of a gas of solar composition in the inner regions of the
protoplanetary disc. This cooling could be a natural consequence of FU Orionis
outbursts in the very innermost regions of the protoplanetary disc
\citep{BellLin1994,HartmannKenyon1996,HillenbrandFindeisen2015} or other
intrinsic long-time-scale variations of the stellar luminosity. The homogeneous
nucleation of iron and silicate particles in a cooling medium was first
considered by \cite{KozasaHasegawa1987} and \cite{KozasaHasegawa1988} in the
context of dust nucleation in stellar outflows and primordial cooling and
condensation in the protoplanetary disc.  They found that large iron nuggets
form when the cooling rate is low. This is due to the high surface tension of
iron \citep{Ozawa+etal2011}. The surface tension determines the stability of the
growing proto-clusters and high surface tension implies (a) that the molecule
has no energetic advantage from nucleating on an existing substrate (i.e.,
heterogeneous nucleation is suppressed) and (b) that the nucleation rates are
low and hence that the rare nano-particles that nucleate from the gas phase will
grow to large sizes by direct depositional growth due to the lack of competition
from other seeds. We extend the classical description of homogeneous nucleation
from \cite{KozasaHasegawa1987} with updated versions of classical nucleation
theory that are better calibrated with experiments even at high values of the
super-saturation level \citep{LummenKraska2005,Kashchiev2006}.

The surface tension of silicates is much lower than for iron
\citep{KozasaHasegawa1987,Zhang+etal2021} and silicates therefore nucleate at a
higher abundance, with correspondingly smaller final sizes.  Iron may deposit as
oxidized FeO on existing silicate particles for high cooling rates
\citep{KozasaHasegawa1988}, but the lowering of the surface tension by admixture
of small amounts of sulfur into the iron causes metallic iron to nucleate at
higher temperatures than FeO. We show that the size difference between iron
pebbles and silicate pebbles leads to the formation of iron-rich planetesimals
by the streaming instability. Mercury and other iron-rich planets could thus be
the result of accretion between such iron-rich planetesimals.

Our paper is organized as follows. In Section 2 we describe our theoretical
model for homogeneous nucleation of iron and a handful of other species. We show
that improved description of classical nucleation theory gives very good fits to
experimental data for homogeneous nucleation of iron. In Section 3 we describe
our model for the protoplanetary disc that contains a zone with active
magnetorotational turbulence in regions where the temperature is above
approximately 800 K. We show that the sublimation fronts of Fe and MgSiO$_3$ are
located close to each other within the active MRI zone. In Section 4 we present
the calculations of the nucleation of several particle sizes and demonstrate how
iron pebbles grow much larger than the pebbles made out of pure silicates. In
Section 5 we discuss implications for the formation of iron-rich planets as well
as shortcomings of the model. We finally conclude with a short summary in
Section 6.

\section{Nucleation model}
\label{s:nucleation_model}

The formation of a solid phase directly from the gas phase is referred to as
homogeneous nucleation, while formation of the solid phase on an existing
substrate is referred to as heterogeneous nucleation \citep{HooseMohler2012}.
We focus first on homogeneous nucleation, since metal and silicate crystal
structures with a high surface tension do not benefit energetically from
nucleating on a substrate material. Later we demonstrate why heterogeneous
nucleation of FeO on existing silicate substrates is likely not a relevant
process in protoplanetary discs.

We summarize here briefly the steps to derive the homogeneous nucleation rate in
classical nucleation theory \citep[see][for a more detailed
picture]{Kashchiev2006}. The change in Gibbs free energy by formation of a solid
nucleus of size $R$ from the gas phase is
\begin{equation}
  \Delta G = \frac{4 \pi}{3} R^3 \Delta g_{\rm v} + 4 \pi R^2 \sigma \, .
\end{equation}
Here $\sigma$ is the surface tension and $\Delta g_{\rm v}$ is the change in
free energy per volume of formation of the solid phase (which has a negative
value).  Small clusters are highly unstable due to the dominant surface tension.
The critical cluster size for stability is found from setting $\partial \Delta
G/\partial R=0$ to obtain
\begin{equation}
  R_{\rm c} = \frac{2 \sigma}{-\Delta g_{\rm v}} \, .
\end{equation}

The free energy difference $\Delta g_{\rm v}$ can now be expressed through the
supersaturation factor $S=P/P_{\rm sat}$ as
\begin{equation}
  \Delta g_{\rm v} = (\rho/\mu) k_{\rm B} T \ln S \, ,
\end{equation}
where $\rho$ is the density of the solid phase and $\mu$ the molecular mass. The
nucleation energy barrier can thus be written as
\begin{equation}
  \Delta G_{\rm c} = \frac{16 \pi \sigma^3}{3 (\Delta g_{\rm v})^2} = \frac{16
  \pi \varOmega^2 \sigma^3}{3 (k_{\rm B} T \ln S)^2} \, .
\end{equation}
Here $\varOmega = \mu / \rho$ is the specific volume per molecule in the solid
phase. The rate at which the cluster of the critical size form is proportional
to the exponential of the ratio of free energy to particle energy,
\begin{equation}
  J = J_0 {\rm e}^{-\Delta G_{\rm c}/(k_{\rm B} T)} \, ,
  \label{eq:Jnuc}
\end{equation}
with the prefactor $J_0$ (of unit ${\rm m^{-3}\,s^{-1}}$). For homogeneous
nucleation, \cite{KozasaHasegawa1987} use the prefactor
\begin{equation}
  J_0 = \alpha_{\rm s} \varOmega \left( \frac{2\sigma}{\pi \mu} \right)^{1/2}
  n_{\rm v}^2 \, .
\end{equation}
Here $\alpha_{\rm s}$ is the sticking probability and $n_{\rm v}$ is the
concentration of vapour monomers in the gas.

\begin{figure}
  \begin{center}
    \includegraphics[width=0.9\linewidth]{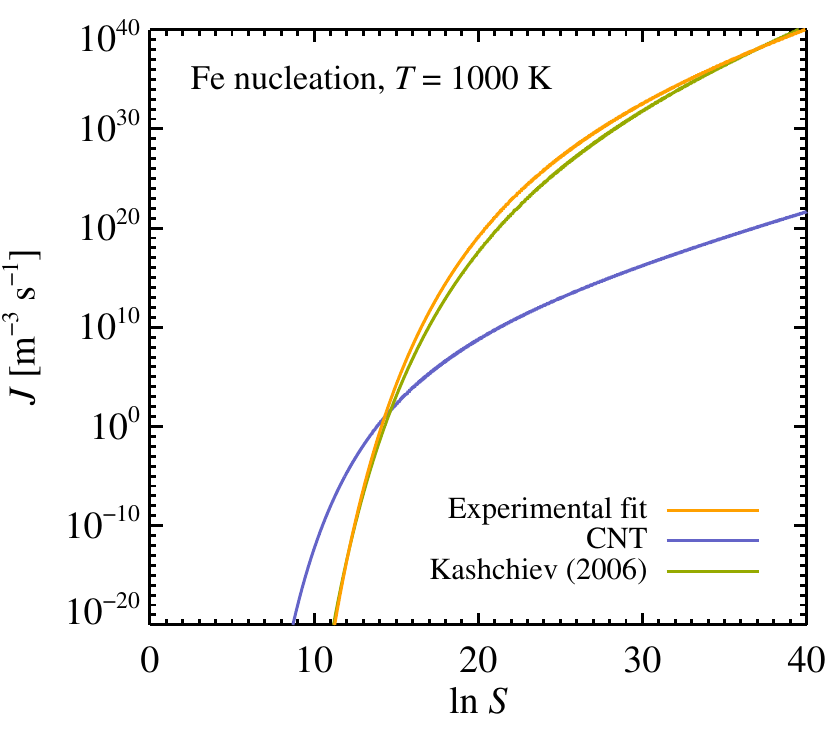}
  \end{center}
  \caption{The homogeneous nucleation rate of iron as a function of the
  logarithmic supersaturation from the experiments of \cite{LummenKraska2005}
  compared to two different nucleation models. The nucleation rate of classical
  nucleation theory (CNT) differs from the experimental results by many orders
  of magnitude. The improved nucleation theory of \cite{Kashchiev2006} fits the
  experimental data much better at both low and high supersaturation.}
  \label{f:homogeneous_nucleation_rate}
\end{figure}

The prefactor $J_0$ is nevertheless highly model-dependent \citep[see][for a
critical review]{Kashchiev2006}. To describe experimental results on homogeneous
nucleation of iron, \cite{LummenKraska2005} use a master equation of the form
\begin{equation}
  J = A_0 S^c \exp \left(-\frac{B}{(\ln S)^2}\right) \, .
  \label{eq:Jgen}
\end{equation}
They determined $A_0$ and $B$ from nucleation experiments at 800 K, 900 K and
1,000 K for different (assumed) values of $c$. For $c=1$ and $T=900$ they found
$B=2.1 \times10^4$ and $\partial \ln B/ \partial \ln T \approx -3.4$. The latter
is consistent with classical nucleation theory, which has a value of -3 (see
equation \ref{eq:Jnuc}), and the value of $B$ is also within order unity equal
to the classical expression. The comparison between data and theory is much
worse for the prefactor scaling $A_0$. We compare the experimental data of
\cite{LummenKraska2005} to the classical nucleation theory expression in Figure
\ref{f:homogeneous_nucleation_rate} and observe that classical nucleation theory
is wrong by many orders of magnitude for both low and high supersaturations.

\cite{Kashchiev2006} therefore adopted corrections to classical nucleation
theory and arrived at the expression
\begin{equation}
  J = \frac{A_0}{16 B^4} \exp \left( \frac{3 B^{1/3}}{4^{1/3}} \right) (\ln
  S)^{12} S \exp \left[ - \frac{B}{(\ln S)^2} \right] \, .
  \label{eq:KNT}
\end{equation}
Here
\begin{equation}
  A_0=\varOmega \left( \frac{2 \sigma}{\pi \mu} \right)^{1/2} n_{\rm s}^2 \, ,
  \label{eq:A0}
\end{equation}
with $n_{\rm s}$ denoting the saturated number density $n_{\rm s} = n_{\rm v} /
S$. The parameter $B$ takes the classical value of
\begin{equation}
  B = \frac{16 \pi \varOmega^2 \sigma^3}{3 (k_{\rm B} T)^3} \, ,
  \label{eq:Bfac}
\end{equation}
but appears in the \cite{Kashchiev2006} model also in the prefactor. We show in
Figure \ref{f:homogeneous_nucleation_rate} the excellent correspondence between
this improved nucleation model and the experimental data on iron nucleation.
Here we multiplied $B$ in equation (\ref{eq:Bfac}) by a small factor 2.5 to
obtain a better fit; this change in $B$ corresponds when propagated to the
nucleation rate $J$ to a sticking probability of approximately 1\% for iron
vapour impinging on the growing iron clusters.

Ironically, despite the orders of magnitude wrong prefactor of classical
nucleation theory for homogeneous nucleation of iron, the value of the
supersaturation where nucleation occurs is not affected very strongly. This can
be seen by first taking the logarithm of the generalized prefactor (equation
\ref{eq:Jgen}),
\begin{equation}
  \ln J - \ln A_0 - c \ln S = - B/(\ln S)^2 \, ,
\end{equation}
and then differentiating while holding the nucleation rate, $\ln J$, constant.
That yields a connection between variations in the prefactor, $\delta \ln A_0$,
and variations in the supersaturation, $\delta \ln S$, as
\begin{equation}
  - \delta \ln A_0 - c \delta \ln S = 2 B \delta \ln S/(\ln S)^3 \, .
\end{equation}
Isolating $\delta \ln S$ gives
\begin{equation}
  \delta \ln S = -\frac{\delta \ln A_0}{2 B/(\ln S)^3 + c} \, .
\end{equation}
Iron generally nucleates in the limit when $B \gg \ln S$. At $T=10^3$ K,
equation (\ref{eq:Bfac}) yields $B \approx 5 \times 10^3$ and hence $\delta \ln
S / \delta \ln A_0 \sim 0.1$ for $\ln S=10$. Hence even an error in $\delta
\ln A_0$ of $\sim$$10$ will be compensated by a $\delta \ln S$ of the order of
1. This relative insensitivity to the prefactor is also evident from Figure
\ref{f:homogeneous_nucleation_rate} where a small shift of the CNT curve along
the $\ln S$ axis leads to an enormous change in the nucleation rate.

\section{Protoplanetary disc model}

We consider the heating and cooling of vapour and solids near the sublimation
temperatures of the refractory molecules Al$_2$O$_3$ (corundum), MgSiO$_3$
(enstatite\footnote{We ignore the major Mg-carrying mineral Mg$_2$SiO$_4$
(forsterite), since it can bind only half of the Si atoms and since its
sublimation temperature is very close to that of enstatite for the pressures
relevant in the inner regions of the protoplanetary disc.}), Fe (iron) and FeO
(w\"ustite). We follow \cite{Nozawa+etal2003}, \cite{Nozawa+etal2011} and
\cite{Woitke+etal2018} and calculate the supersaturation level $S$ from the
chemical reaction that forms the relevant mineral. We include the following
reactions separately,
\begin{eqnarray}
  2\, {\rm Al} ({\rm g}) + 3\, {\rm H}_2{\rm O} ({\rm g}) &\rightleftharpoons&
  {\rm Al_2O_3} ({\rm s}) + 3\, {\rm H_2} ({\rm g}) \, , \\
  {\rm Mg} ({\rm g}) + {\rm SiO} ({\rm g}) + 2\, {\rm H}_2{\rm O} ({\rm
  g})&\rightleftharpoons& {\rm MgSiO_3} ({\rm s}) + 2\, {\rm H_2} ({\rm g}) \, ,
  \\
  {\rm Fe} ({\rm g}) &\rightleftharpoons& {\rm Fe} ({\rm s})\, ,  \\
  {\rm Fe} ({\rm g}) + {\rm H}_2{\rm O} ({\rm g}) &\rightleftharpoons& {\rm
  FeO} ({\rm s}) + {\rm H_2} ({\rm g}) \, , \\
  {\rm H_2O} ({\rm g}) &\rightleftharpoons& {\rm H_2O} ({\rm s})\, .
\end{eqnarray}
As an example, the supersaturation level of Al$_2$O$_3$ (corundum) is calculated
from the equilibrium
\begin{eqnarray}
  \ln S &=& - \Delta G / (R_{\rm gas} T) + 2 \ln(P_{\rm Al}/P_{\rm std}) + 3 \ln
  (P_{\rm H_2O}/P_{\rm std}) \nonumber \\ && - 3 \ln (P_{\rm H_2}/P_{\rm std})
  \, ,
\end{eqnarray}
where $P_{\rm std}=10^5$ Pa is the pressure of the standard state and $R_{\rm
gas}$ is the universal gas constant. Here the chemical activity of the mineral
phase (Al$_2$O$_3$) is set to zero \citep{Nozawa+etal2003}. The change in Gibbs
free energy of this reaction, $\Delta G$ (per mole), is calculated by
subtracting the sum of the Gibbs free energies of the reactants from the sum of
the Gibbs free energies of the products (including the solid mineral), following
the parametrizations presented in \cite{SharpHuebner1990}. This chemical
reaction solution to the formation of a mineral phase corresponds to a saturated
vapour pressure approach only when the mineral formation does not involve a
chemical reaction with another species, as is the case above for Fe. Using a
saturated vapour pressure approach to the more complex minerals on our list
would only give an approximate estimate of the equilibrium temperature
\citep{vanLieshout+etal2014,Aguichine+etal2020}.

We consider the inner regions of the protoplanetary disc and construct the
temperature ($T$) and surface density ($\varSigma$) profiles as function of the
distance $r$ using an assumed constant gas accretion rate $\dot{M}_\star$
through the disc and onto the star of mass $M_\star$ and luminosity $L_\star$.
The temperature is taken as the maximum of the irradiative temperature
\citep{Ida+etal2016},
\begin{equation}
  T_{\rm irr} = 150\,{\rm K}\,\left(\frac{L_\star}{L_\odot}\right)^{2/7}
  \left(\frac{M_\star}{M_\odot}\right)^{-1/7} \left( \frac{r}{\rm AU}
  \right)^{-3/7} \, ,
\end{equation}
and the viscous temperature $T_{\rm vis}$. The latter is found from the balance
between viscous heating and black body cooling at the disc surface
\citep{BellLin1994},
\begin{equation}
  \frac{3}{4 \pi} \dot{M}_\star (\delta/\alpha) \varOmega_{\rm
  K}^2 = 2 \sigma_{\rm SB} T_{\rm eff}^4 \, .
\end{equation}
Here $T_{\rm eff}$ is the effective cooling temperature at the disc surface.  We
take into account that a part of the mass accretion rate will not lead to
heating, if the angular momentum is transported by large-scale disc winds
\citep{Mori+etal2021}. Hence $\delta/\alpha$, with $\alpha$ representing the
total angular transport and $\delta$ representing the local turbulence,
gives the fraction of mass transported via turbulence and hence leading to
heating.
\begin{figure}
  \begin{center}
    \includegraphics[width=\linewidth]{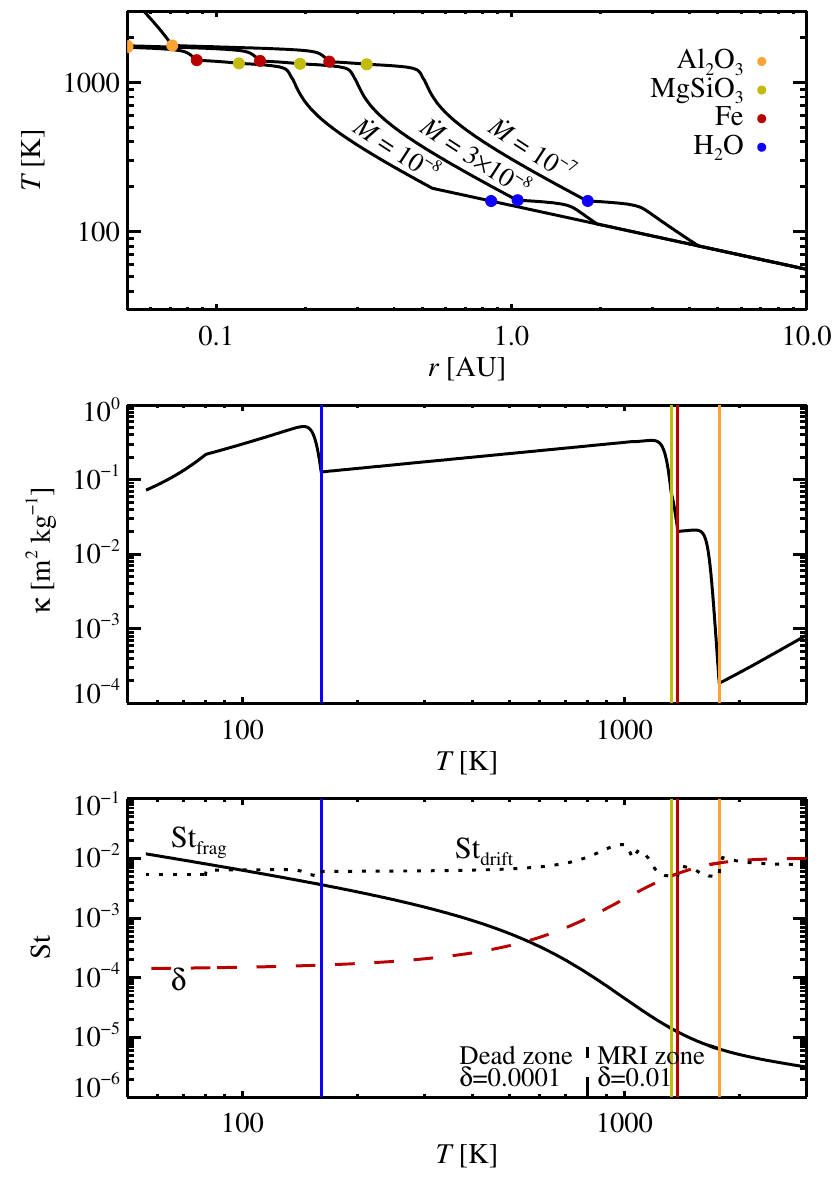}
  \end{center}
  \caption{Thermal structure of the inner regions of a protoplanetary disc at
  three different stellar mass accretion rates (top panel), total gas and dust
  opacity as a function of temperature (middle panel) and particle Stokes number
  and diffusion coefficient as a function of temperature (bottom panel, at
  $\dot{M}=10^{-7}M_\odot\,{\rm yr}^{-1}$). The sublimation fronts of corundum
  (Al$_2$O$_3$), enstatite (MgSiO$_3$), iron (Fe) and water (H$_2$O) are
  indicated. The fragmentation-limited Stokes number is much lower than the
  drifting Stokes number interior of the water ice line and hence the radial
  motion of solids is dominated by the gas accretion speed in the inner regions
  of the protoplanetary disc.}
  \label{f:sublimation_fronts}
\end{figure}
The effective temperature is connected to the mid-plane temperature $T_{\rm
vis}$ through radiative balance
\begin{equation}
  T_{\rm vis}^4 = \frac{3}{8} \kappa \varSigma T_{\rm eff}^4 \, .
\end{equation}
Here $\kappa$ is the opacity. The opacity is found by calculating the mass
fraction in solids from the saturated vapour densities $\rho_{{\rm sat},i}$ as
\begin{equation}
  \rho_{{\rm sol},i} = {\rm min}(\rho_{{\rm sat},i},\rho_i)
\end{equation}
We then add up the opacity power laws from \cite{BellLin1994} for water ice,
``metals'' (iron, enstatite and corundum) and gas, with each component
multiplied by its relative depletion factor due to thermal desorption.

We assume that the turbulent diffusion varies from $\delta_{\rm DZ}=10^{-4}$ in
the dead zone to $\delta_{\rm MRI}=10^{-2}$ where the magnetorotational
instability is active, following the profile
\begin{eqnarray}
  \log_{10} \delta &=& \log_{10} \delta_{\rm DZ} + (\log_{10} \delta_{\rm MRI} -
  \log_{10} \delta_{\rm DZ} ) \nonumber \\ &\times& \left[ \tanh \left(
  \frac{T-T_{\rm MRI}}{W_{\rm MRI}} \right) +1 \right] /2 \, .
\end{eqnarray}
We use $T_{\rm MRI}=800$ K \citep{DeschTurner2015} and $W_{\rm MRI}=600$ K
\citep[which gives a good match to the turbulence profile in the inner regions
of the protoplanetary disc, see][]{Flock+etal2017}.

We solve the structure equations iteratively to find $T(r)$ and $\varSigma(r)$.
The resulting temperature profile is shown in Figure \ref{f:sublimation_fronts}
for three different values of the mass accretion rate onto the star. The
sublimation fronts of iron and enstatite are very close together and evolve from
0.2--0.3 AU at high accretion rates to 0.1 AU for the lower accretion rates
characteristic of later disc phases.
\begin{figure}
  \begin{center}
    \includegraphics[width=0.9\linewidth]{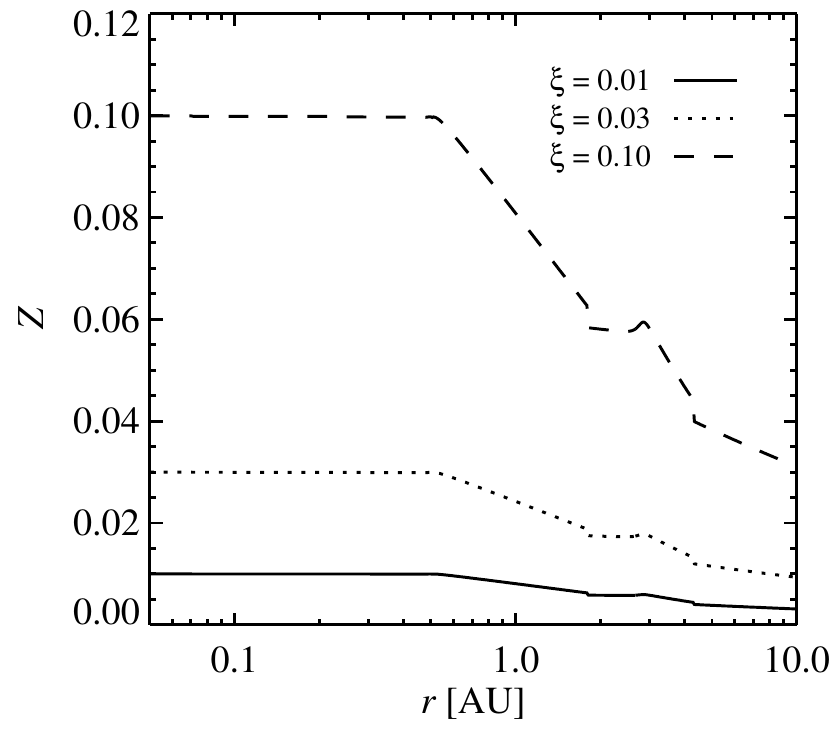}
    \includegraphics[width=0.9\linewidth]{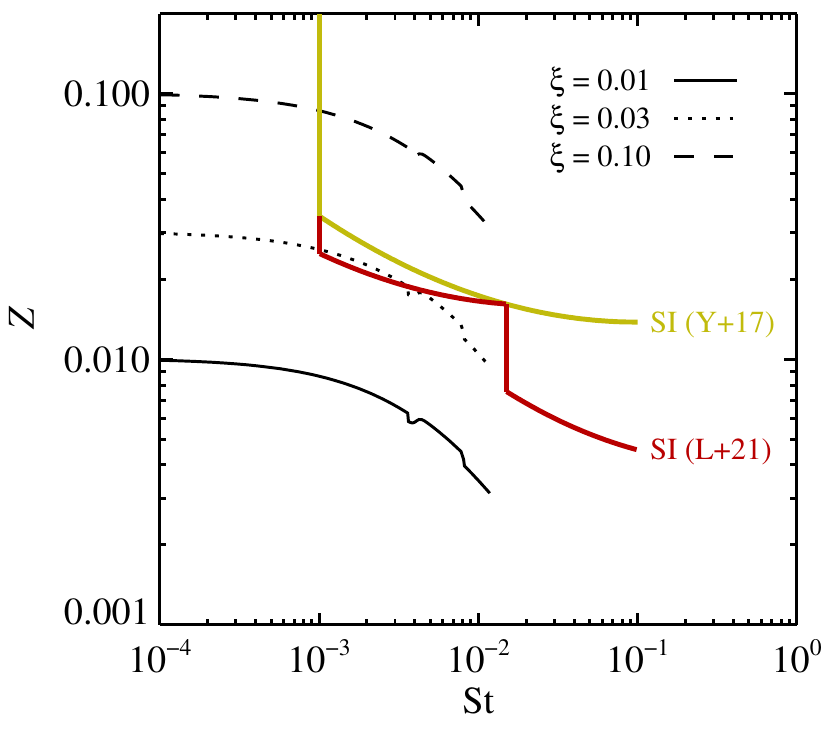}
  \end{center}
  \caption{The mass loading of solid particles in the gas as a function of
  distance from the star (top plot) and the mass loading as a function of the
  Stokes number (bottom plot). Results are shown for $\dot{M}_\star = 10^{-7}
  M_\odot\,{\rm yr}^{-1}$ and three different values of the pebble-to-gas flux
  ratio $\xi=\dot{M}_{\rm p}/\dot{M}_\star$. We overplot the threshold
  metallicity for planetesimal formation by the streaming instability in the
  bottom plot, from two different sources \citep{Yang+etal2017,LiYoudin2021}. An
  unrealistically large pebble flux of $\xi>0.03$ is needed to trigger the
  streaming instability directly for the considered values of ${\rm St}$.}
  \label{f:metallicity}
\end{figure}

The fragmentation-limited Stokes number is calculated from
\citep{Birnstiel+etal2012}
\begin{equation}
  {\rm St}_{\rm frag} = \frac{1}{3} \frac{1}{\delta} \left( \frac{v_{\rm
  frag}}{c_{\rm s}} \right)^2 \, ,
\end{equation}
where $v_{\rm frag}=1$ m/s is the fragmentation threshold speed
\citep{Guettler+etal2010} and $c_{\rm s}$ is the isothermal sound speed of the
gas defined as $c_{\rm s} = \sqrt{k_{\rm B} T/\mu}$ with $\mu$ representing
the mean molecular weight of a solar mixture of H$_2$ and He. Here the
dimensionless Stokes number is defined as
\begin{equation}
  {\rm St} = \sqrt{2 \pi} \frac{R \rho_\bullet}{\varSigma} \, ,
\end{equation}
where $R$ is the particle radius, $\rho_\bullet$ is the internal density and
$\varSigma$ is the gas surface density. This definition of the Stokes
number is valid in the Epstein regime when the particle size is much smaller
than the mean free path of the gas \citep{Weidenschilling1977}. The particles
drift towards the star with the speed \begin{equation}
  v_r = - 2 {\rm St} \Delta v + u_r \, .
\end{equation}
This expression is valid for ${\rm St} \ll 1$
\citep{Weidenschilling1977,Johansen+etal2019}.  Here the gas speed is given by
\citep{Pringle1981}
\begin{equation}
  u_r = - \frac{3}{2} \frac{\nu}{r} - 3 \frac{\partial (\varSigma \nu)/\partial
  r}{\varSigma} \, .
\end{equation}
For very low Stokes numbers, $v_r \approx u_r$, the particles are primarily
transported by gas advection. We define the Stokes number where the radial drift
speed is as high as the gas advection speed from $2 {\rm St}_{\rm drift} \Delta
v = u_r$.

Figure \ref{f:sublimation_fronts} shows that at temperatures below the ice
sublimation temperature, the fragmentation limited Stokes number is large enough
that particles drift faster than the gas. In that case the pebble-to-gas flux
ratio can become higher than the nominal value of $\xi=0.01$
\citep{Ida+etal2016,JohansenBitsch2019}. Here $\xi$ is defined as the ratio of
the radial pebble flux to the radial gas flux,
\begin{equation}
  \xi = \frac{\dot{M}_{\rm p}}{\dot{M}_\star} \, .
\end{equation}
Knowing $\xi$, we can reconstruct the local metallicity from the relationship
\begin{equation}
  Z = \frac{\varSigma_{\rm p}}{\varSigma_{\rm g}} = \frac{\dot{M}_{\rm
  p}}{\dot{M}_\star} \frac{u_r}{v_r} \, .
\end{equation}
We show the metallicity profile for three different values of $\xi$ in Figure
\ref{f:metallicity}. In the inner regions of the protoplanetary disc, where the
Stokes number is low and $v_r \approx u_r$, we recover $Z=\xi$. Further out,
$\xi$ drops. We also show in Figure \ref{f:metallicity} the threshold
metallicity for planetesimal formation by the streaming instability. It is clear
that very high pebble-to-gas flux ratios, of $\xi>0.03$, are needed to trigger
the streaming instability for Stokes numbers in the range between $10^{-3}$ and
$10^{-2}$, which may be achievable by fragmentation-limited growth in the dead
zone of the protoplanetary disc. The much lower Stokes number in the MRI zone
are not prone to planetesimal formation by the streaming instability for any
values of $\xi$.. Larger particle sizes are needed to form planetesimals in the
region where Mercury formed -- and we propose here that these particles form by
nucleation and depositional growth.

\section{Cooling, nucleation and phase separation}

Young stars undergo periodic outbursts where luminosities may reach
$10^2-10^3\,L_\odot$ and mass accretion rates onto the star are as high as
$10^{-4}\,M_\odot\,{\rm yr^{-1}}$ \citep{HartmannKenyon1996}. These outbursts
are likely related to a thermal instability that affects the very innermost
regions of the disc where the effective temperature of the disc is higher than
$T_{\rm eff} = 2000$ K. \cite{BellLin1994} estimated that the ionization front,
which separates the outburst region of the protoplanetary disc from the
quiescent part, will move out as far as 0.1 AU from the star. This means that
regions of iron and silicate sublimation are not directly involved in the
outbursts. However, the strong irradiative heating at the outburst luminosity
will, during the outburst, affect even the Mercury formation region, with the
outburst temperature reaching
\begin{equation}
  T_{\rm out} = 1600\,{\rm K}\,\left(\frac{L_\star}{10^3L_\odot}\right)^{2/7}
  \left(\frac{M_\star}{M_\odot}\right)^{-1/7} \left( \frac{r}{0.4\,{\rm AU}}
  \right)^{-3/7} \, .
\end{equation}
These FU Ori outbursts decay on time-scale of decades or even centuries
\citep{HartmannKenyon1996}. Other models that rely on the triggering of MRI
turbulence and gravitational instability have FU Orionis outbursts extended out
to 0.5 AU or more
\citep{Zhu+etal2007,Zhu+etal2009,HillenbrandFindeisen2015,Pignatale+etal2018}.
In that case, the nucleation of refractory dust can also take place due to the
decrease in mass accretion rate and local viscous heating as the outburst
declines.
\begin{figure}
  \begin{center}
    \includegraphics[width=0.9\linewidth]{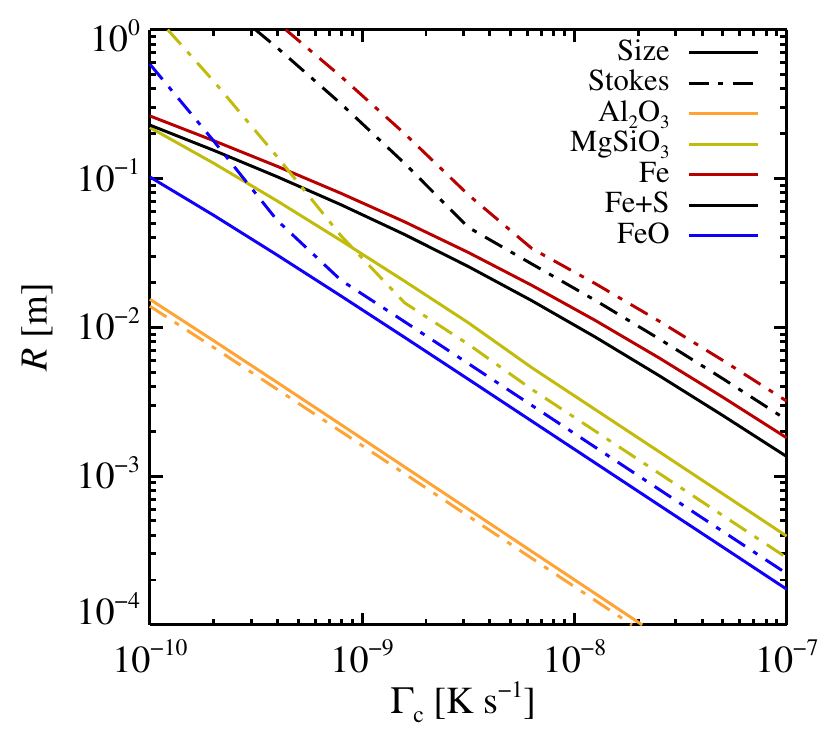}
    \includegraphics[width=0.9\linewidth]{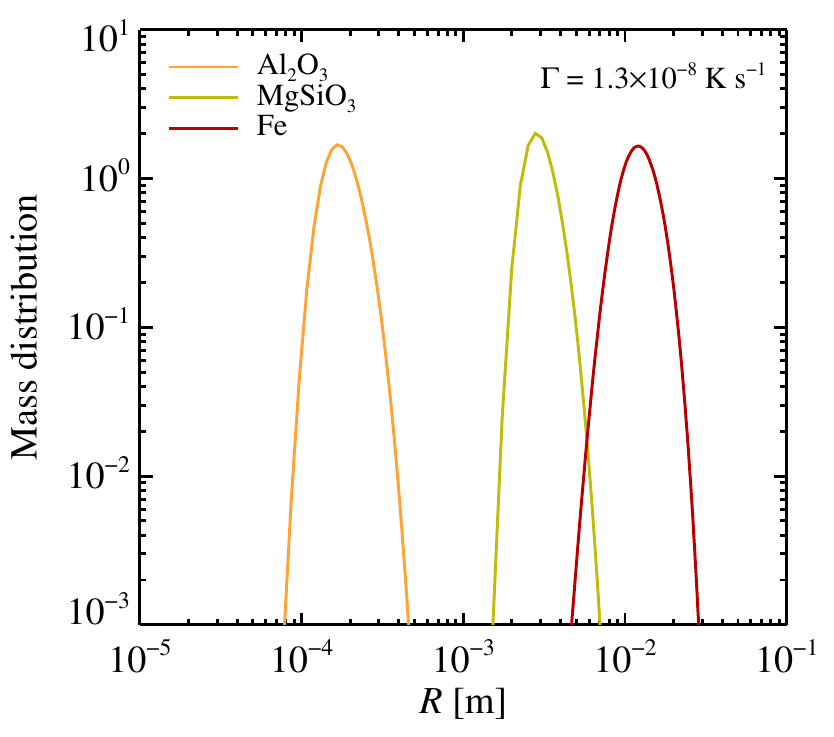}
  \end{center}
  \caption{Particle sizes and Stokes numbers as a function of the cooling rate
  of our protoplanetary disc model at $\dot{M}=10^{-7}\,M_\odot\,{\rm yr}^{-1}$
  at $r=0.4$ AU (top panel) and size distribution of corundum, enstatite and
  iron at a specific cooling rate (bottom panel). All the species increase in
  size for decreasing cooling rates. The Stokes number increases
  quadratically with the particle size above the size transition from
  Epstein to Stokes regime. Pure iron (Fe) and iron with sulfur (Fe+S) obtain
  the largest sizes and, by far, the highest Stokes numbers.}
  \label{f:homogeneous_nucleation}
\end{figure}
\begin{figure}
  \begin{center}
    \includegraphics[width=0.9\linewidth]{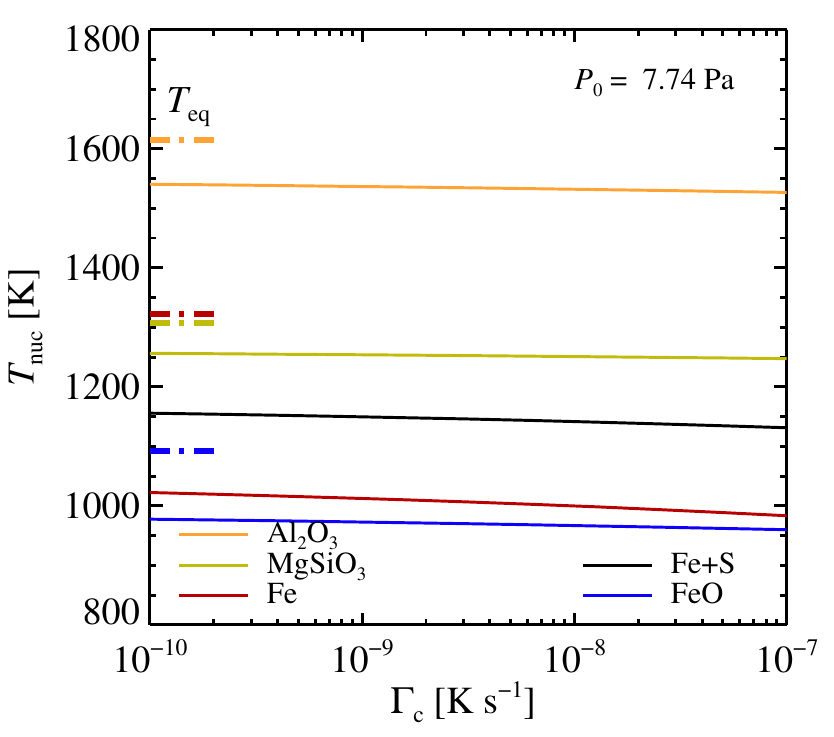}
    \includegraphics[width=0.9\linewidth]{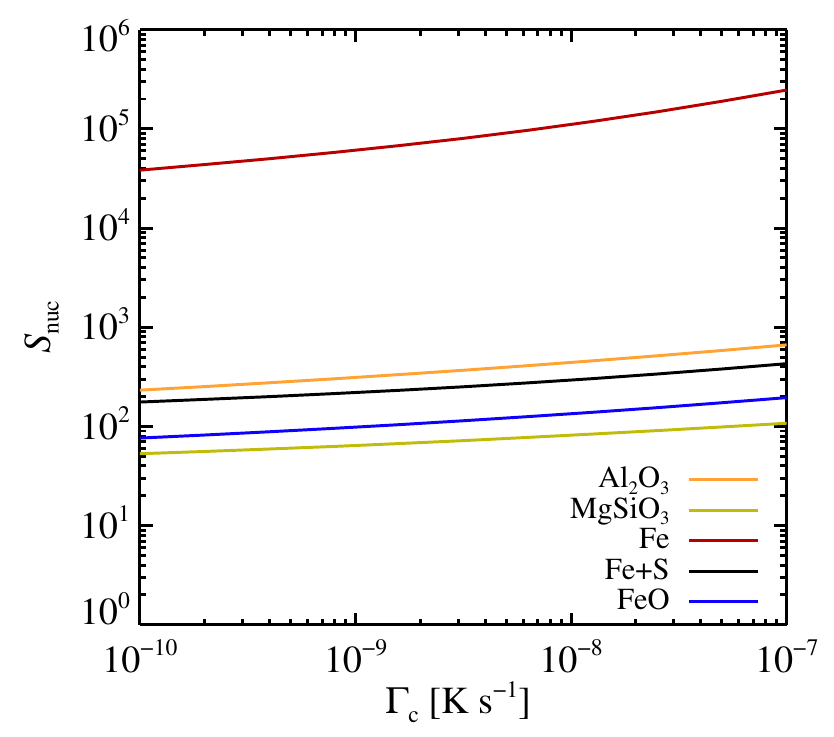}
  \end{center}
  \caption{The nucleation temperature (top panel) and nucleation saturation at
  maximum nucleation rate (bottom panel) as a function of the cooling time.  The
  equilibrium temperatures, where the gas phase is in balance with the solid
  phase, are indicated with short dot-dashed lines. All the species nucleate at
  temperatures significantly below their equilibrium temperatures, due to the
  inhibition of nucleation at low super-saturation levels. Iron in its pure form
  nucleates at lower temperature than the equilibrium of FeO, which would lead
  instead to direct condensation of FeO onto existing silicates. However, the
  presence of small amounts of sulfur in the iron (Fe+S) lowers the surface
  tension enough that iron nucleates before the condensation of FeO.}
  \label{f:nucleation_temperature}
\end{figure}

The cooling rate of an outburst can be connected to the decay time-scale
$\tau_\star = L_\star / \dot{L}_\star$ as
\begin{eqnarray}
  \varGamma_{\rm c} \equiv - \dot{T} &=& 1.5 \times 10^{-8}\,{\rm K\,s^{-1}}\,
  \left( \frac{L_\star/\dot{L}_\star}{10^3\,{\rm yr}} \right)^{-1}
  \left(\frac{L_\star}{10^3L_\odot}\right)^{2/7} \nonumber \\ & & \times
  \left(\frac{M_\star}{M_\odot}\right)^{-1/7} \left( \frac{r}{0.4\,{\rm AU}}
  \right)^{-3/7}  \, .
\end{eqnarray}
Hence realistic cooling rates in the Mercury-formation zone lie in the range of
$\varGamma_{\rm c}=10^{-8}-10^{-7}\,{\rm K\,s^{-1}}$.

We calculate the nucleation of particles at $0.4$ AU, the approximate current
location of Mercury, as a function of the cooling rate $\varGamma_{\rm c}$. The
nucleation rate is calculated from equation (\ref{eq:KNT}). Nucleated particles
grow at the rate determined by the flux of the constituent vapour species with
the lowest impact frequency, the so-called key vapour species
\citep{Nozawa+etal2003}, according to the equation
\begin{equation}
  \dot{R} = \frac{\alpha_{\rm s} \alpha_{\rm d}}{\alpha_{\rm s} + \alpha_{\rm
  d}} \frac{n_{\rm v}}{\rho_\bullet/\mu_\bullet} v_\perp \, ,
\end{equation}
where $v_\perp = \sqrt{k_{\rm B} T/(2 \pi m_{\rm v})}$ is the speed of the key
vapour species perpendicular to the surface, $n_{\rm v}$ is the number density
of the key vapour species, $\rho_\bullet$ and $\mu_\bullet$ are the density and
molecular weight of the solid phase, $\alpha_{\rm s}$ is the sticking
probability and $\alpha_{\rm d} = D/(R v_\perp)$ is a correction factor for
large grains whose growth is limited by the diffusion coefficient $D$
\citep{Libbrecht2005}. We set the sticking coefficient $\alpha_{\rm s}$ of Fe
vapour onto the Fe solid phase to be unity due to the highly super-saturated
vapor deposition \citep{Tachibana+etal2011}. We set the sticking coefficient to
0.1 for all other species due to kinetic hindrance of evaporation and
condensation \citep{Takigawa+etal2009,vanLieshout+etal2014}.

Following \cite{KozasaHasegawa1988} we include the species Al$_2$O$_3$,
MgSiO$_3$, Fe, Fe+S and FeO. We added here Fe+S to our list in order to probe
the effect of sulfur on the nucleation rate of iron, since dissolved sulfur
lowers the surface tension of iron \citep{KozasaHasegawa1988} and hence
increases the nucleation temperature to closer to the equilibrium temperature.
FeO plays the additional role to probe whether iron nucleates at higher or lower
temperatures than the equilibrium temperature of FeO over a silicate surface
(which would mix iron and silicates).
\begin{figure}
  \begin{center}
    \includegraphics[width=0.9\linewidth]{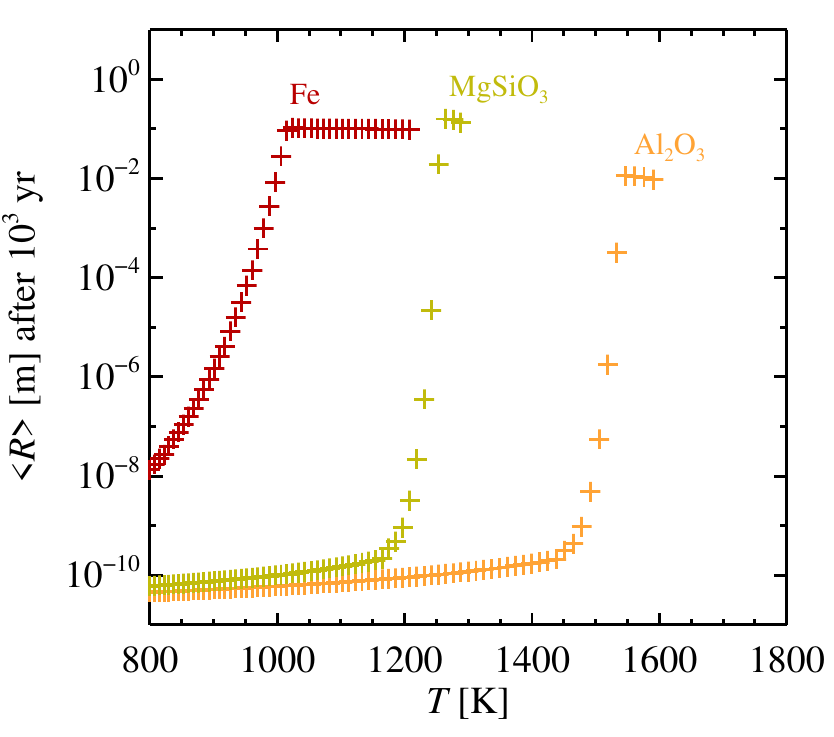}
  \end{center}
  \caption{The mean particle size of Fe, MgSiO$_3$ and Al$_2$O$_3$ after $10^3$
  yr exposed to a range of temperatures. Nucleation is so inefficient at large
  temperatures that meter-sized particles grow by vapour deposition on the very
  rare nuclei, but these particles do not absorb most of the vapour mass, as
  some vapour will not have time to undergo deposition. Lowering the temperature
  leads to a sharp decline in the particle sizes (and the fraction of vapour
  deposited on the grains increases abruptly from 0\% to 100\% below the
  transition temperature).}
  \label{f:homogeneous_nucleation_nocool}
\end{figure}

The surface tension is a key parameter that determines the outcome of nucleation
through equations (\ref{eq:KNT}) and (\ref{eq:A0}). We use nominal values from
\cite{KozasaHasegawa1987} and \cite{KozasaHasegawa1988} of $\sigma=0.69\,{\rm
J\,m^{-2}}$ for corundum, $\sigma=0.4\,{\rm J\,m^{-2}}$ for enstatite,
$\sigma=1.8\,{\rm J\,m^{-2}}$ for iron and $\sigma=0.6\,{\rm J\,m^{-2}}$ for
FeO. The effect of sulfur in reducing the surface tension of iron is a
strongly declining function of temperature. As we show below, this surface
tension reduction is key to prevent heterogeneous nucleation of iron oxide (FeO)
on existing enstatite substrates. We assume that the presence of S in the Fe
crystals lowers the surface tension of Fe+S to $\sigma=1.2\,{\rm J\,m^{-2}}$,
which we read off from Figure 2 of \cite{KozasaHasegawa1988} at a temperature of
approximately 1150 K. The strong temperature dependence of the sulfur activity
implies that any difference in the stellar S/Fe ratio relative to the solar
value is likely not significant when calculating the surface tension.

The resulting particle sizes and Stokes numbers are shown in Figure
\ref{f:homogeneous_nucleation}. We take into account here that large
particles enter the Stokes drag force regime where the Stokes number St scales
quadratically with the particle size \citep{Weidenschilling1977}. There is a
general development towards larger particle sizes for lower cooling rate. This
is due to the increased time for condensation onto rare nuclei when the cooling
rate is low.  Higher cooling rates lead to the nucleation of many more
nano-particles and hence the resulting particle size becomes smaller. The
largest particles are Fe (and Fe+S) due to the very high surface tension, and
hence limited nucleation, of iron. These iron particles are generally five times
larger than silicate particles (MgSiO$_3$) and have ten times higher Stokes
number.

The nucleation temperatures and supersaturation levels at the maximum nucleation
are shown in Figure \ref{f:nucleation_temperature}. The main nucleation
occurs several hundred Kelvin below the formal equilibrium temperatures; this is
true for all species. Iron has so poor nucleation properties that the nucleation
temperature is even below the equilibrium temperature of FeO. Hence iron atoms
would instead deposit as FeO onto existing silicates. However, inclusion of
sulfur in the iron (Fe+S) decreases the surface tension enough that Fe+S can
nucleate before FeO and hence prevent the iron from getting oxidized.
\begin{figure*}
  \begin{center}
    \includegraphics[width=0.9\linewidth]{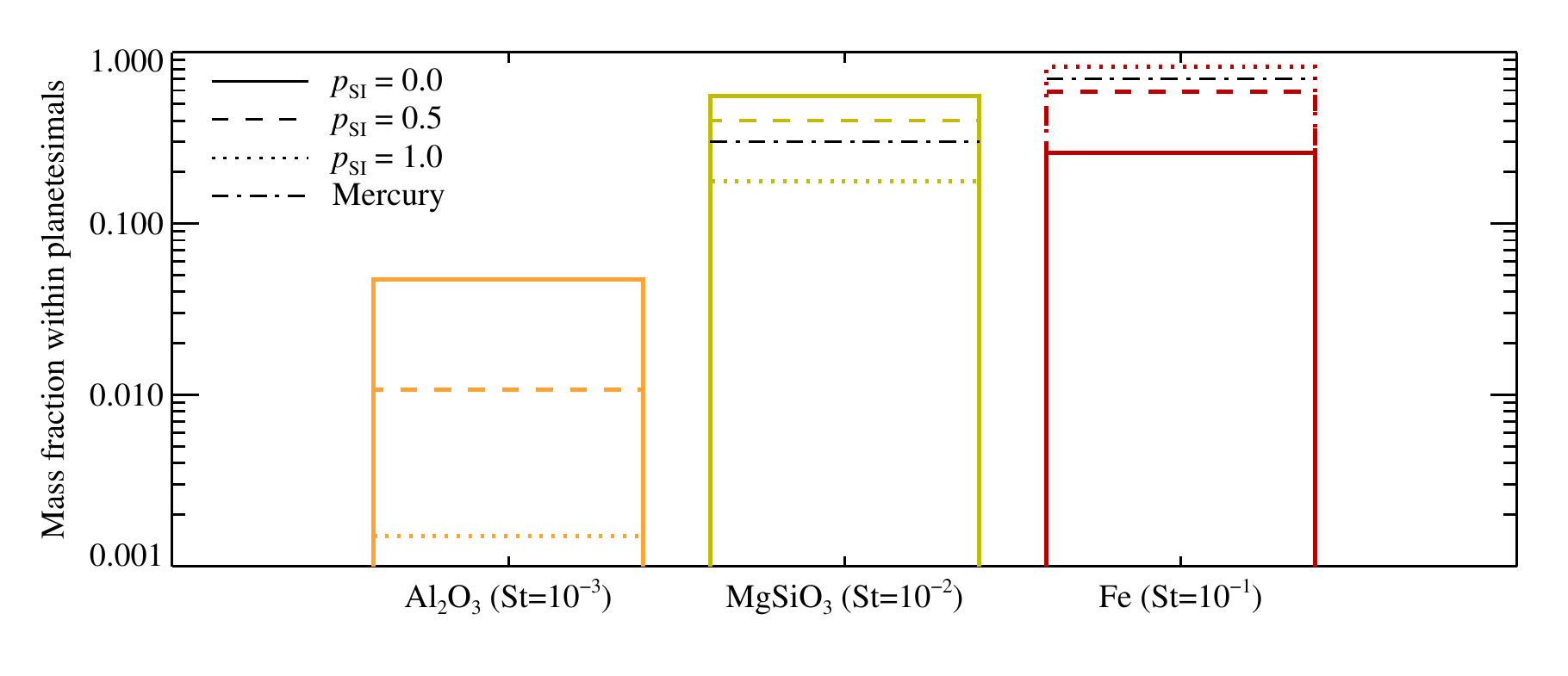}
  \end{center}
  \caption{Mass distribution within planetesimals formed from streaming
  instability filaments, compared to the composition of Mercury. We assume here
  that iron pebbles have ${\rm St}=0.1$, enstatite pebbles have ${\rm St}=0.01$
  and corundum pebbles have ${\rm St}=0.001$. We consider filtering function
  power laws ranging from $p_{\rm SI}=0$ (where particles maintain their size
  distribution when they enter planetesimals) to $p_{\rm SI}=1.0$ (where smaller
  particles have a reduced presence in planetesimals proportional to their
  Stokes number). We find a good match to Mercury between $p_{\rm SI}=0.5$ and
  $p_{\rm SI}=1.0$. This filtering additionally implies that Mercury will be
  strongly depleted in small corundum particles. This is consistent with
  the measured low Al/Si ratio in Mercury's crust \citep{Peplowski+etal2012}.}
  \label{f:relative_mass_filaments}
\end{figure*}

We perform an additional numerical experiment where we instantaneously lower the
temperature from 2,000 K to a (fixed) target temperature and let the particles
nucleate and grow at a constant temperature. The resulting particle sizes are
shown in Figure \ref{f:homogeneous_nucleation_nocool}. At large temperatures,
nucleation is so inefficient that the rare nuclei can grow to meter sizes by
direct vapour deposition. This is nevertheless a very inefficient process as not
all the vapour will be transferred to the solid phase even in the $10^3$ yr
integration of the experiment. At lower temperatures, vapour deposition leads to
smaller particles, but the efficiency over $10^3$ yr is here 100\%
(i.e., all the vapour is deposited onto the particles). As the temperature is
pushed 100 K or more below this transition to efficient deposition, the
nucleation becomes so efficient that all vapour resides in nano-meter-scale
particles.

\section{Discussion}

\subsection{Formation of iron-rich planetesimals}

The density and gravity field of Mercury yield a bulk composition of between
60\% and 70\% iron \citep{Hauck+etal2013}. This is approximately twice the
iron-to-rock ratio of the solar composition \citep{Lodders2003}. The highly
reduced mantle of Mercury furthermore implies that the planet could have
incorporated a significant fraction of Si into the metallic melt during core
formation \citep{Nittler+etal2018}, but this does not change the overall picture
that Mercury has an enlarged iron fraction relative to the other terrestrial
planets.

We consider now the implications of our dust nucleation model for planetesimal
formation by the streaming instability. The pebbles with the highest Stokes
numbers are known to drive filament formation by the streaming instability when
considering a size distribution of the pebbles
\citep{BaiStone2010,Schaffer+etal2018,Schaffer+etal2021}. The size distribution
of the pebbles entering the planetesimals that condense out of the filament is
nevertheless relatively unexplored. We define here a ``filtering function''
$P({\rm St})$ that describes the amount of pebbles of a certain Stokes number
that enter a planetesimal divided by the total number of pebbles of that Stokes
number,
\begin{equation}
  P({\rm St}) = ({\rm St}/{\rm St}_{\rm max})^{p_{\rm SI}} \, .
\end{equation}
Here ${\rm St}_{\rm max}$ is the highest Stokes number in the size distribution.
Thus the differential of the cumulative mass distribution $F_>({\rm St})$ (i.e.,
the fraction of mass present in pebbles with Stokes number larger than St) in
the planetesimal becomes
\begin{equation}
  \frac{{\rm d} F_>}{{\rm d} {\rm St}} \bigg\rvert_{\rm pla} = P({\rm St})
  \frac{{\rm d} F_>}{{\rm d} {\rm St}} \bigg\rvert_{\rm disc}
\end{equation}
From the results of \cite{Johansen+etal2007} we estimate $p \approx 1.0$ for the
two Stokes number bins with ${\rm St}=0.25$ and ${\rm St}=0.50$; the bins with
even higher Stokes numbers contribute to the planetesimals relatively evenly but
this may reflect that such large pebbles are already quite decoupled from the
gas. From the non-stratified simulations of \cite{YangZhu2021} we find instead a
steeper $p_{\rm SI} \approx 1.5$ in the densest clumps when considering Stokes
numbers in the range between $0.01$ and $0.05$ that are more relevant for our
nucleation model. We show in Figure \ref{f:relative_mass_filaments} how values
of $p_{\rm SI}$ between 0.5 and 1.0 match Mercury well when there is a factor
ten difference in Stokes number between iron and enstatite pebbles. Mercury's
crust also appears to have an enriched Al/Si ratio relative to the moon
\citep{Peplowski+etal2012}; this could indicate either an affinity with
enstatite chondrites (which are rich in Si) or a limited accretion of
Al-carrying minerals such as corundum (Al$_2$O$_3$).

\subsection{Formation of Mercury analogues and super-Mercuries}

The formation of iron-rich planetesimals will be followed by growth of
protoplanets through planetesimal accretion and pebble accretion. The growth
will initially take place in the MRI-active zone of the protoplanetary disc and
therefore excitation of the orbits of the protoplanets by turbulent density
fluctuations must be take into account. The eccentricity induced by the
turbulent fluctuations in balance with gas drag is
\citep{Ida+etal2008,JohansenBitsch2019}
\begin{eqnarray}
  e_{\rm drag} &=& 0.075 f_{\rm g}^{1/3} \left( \frac{\gamma}{10^{-2}}
  \right)^{2/3} \left( \frac{R}{2.4 \times 10^6\,{\rm m}} \right)^{1/3}
  \nonumber \\ && \qquad \qquad \times \left( \frac{\rho_\bullet}{5.4 \times
  10^3\,{\rm kg\,m^{-3}}}\right)^{1/3} \left( \frac{r}{0.4\,{\rm AU}}
  \right)^{11/12} \, .  \label{eq:edrag}
\end{eqnarray}
Here $\gamma$ is a measure of the strength of the turbulent density fluctuations
and we assume $\gamma \sim \alpha$ as in \cite{Ida+etal2008}. The parameter
$f_{\rm g}$ describes the gas density relative to the minimum mass solar nebula.
We scaled the equations to the radius and orbit of Mercury. Tidal damping is
relevant for large protoplanets and planets. \cite{Ida+etal2008} give the
equilibrium eccentricity between turbulent density fluctuations and tidal
damping as
\begin{eqnarray}
  e_{\rm tidal} &=& 0.024 f_{\rm g}^{1/2} \left( \frac{\gamma}{10^{-2}}
  \right)^{1/2} \left( \frac{R}{2.4 \times 10^6\,{\rm m}} \right)^{1/3} \nonumber \\
  && \qquad \qquad \times \left( \frac{\rho_\bullet}{5.4 \times 10^3\,{\rm
  kg\,m^{-3}}}\right)^{-1/2} \left( \frac{r}{0.4\,{\rm AU}} \right)^{3/4} \, .
  \label{eq:etidal}
\end{eqnarray}
Mercury's current eccentricity of 0.2 is therefore hard to excite by realistic
values for the turbulent stirring and likely reflects later perturbations by
Jupiter \citep{Roig+etal2016}. As the MRI-zone retreats with time and the
protoplanets enter the dead zone, the eccentricity is reduced by 1-2 orders of
magnitude and pebble accretion becomes the main growth mode.

Mercury must have avoided a phase of rapid pebble accretion in order to have
maintained its low mass and to prevent dilution of its iron-rich
composition with drifting pebbles of solar Fe/Si ratio. It is possible that the
inner regions of the protoplanetary disc were already dissipated away by disc
winds when the MRI zone retreated to interior of Mercury's orbit
\citep{Ogihara+etal2017}. Alternatively, the inner disc edge had expanded
outwards across Mercury's orbit when the temperature became cold enough for
pebble accretion to be efficient \citep{Liu+etal2017}. The oblateness of
the young Sun could also have increased Mercury's inclination to its current
value if the Sun had a high primordial rotation rate \citep{Ward+etal1976} and
thus suppressed pebble accretion.

\subsection{Growth of planetesimals}

\begin{figure}
  \includegraphics[width=0.9\linewidth]{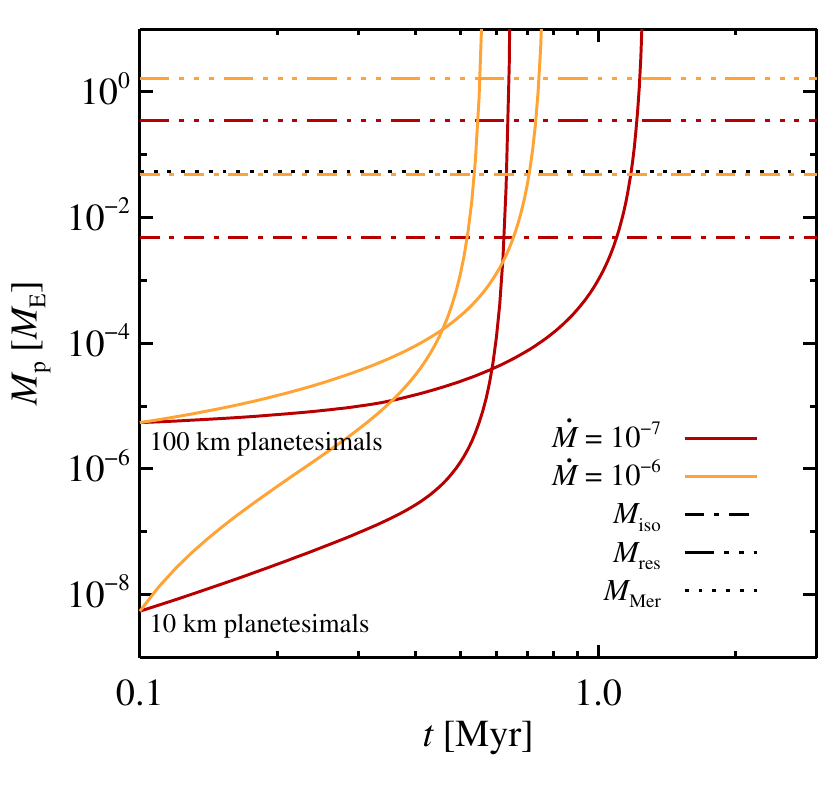}
  \caption{The growth of a Mercury analogue at $r=0.4$ AU by accretion of
  iron-rich planetesimals. We consider planetesimal sizes of either 100 km
  (nominal) or 10 km (small planetesimals) and two values for the accretion rate
  of the protoplanetary disc when the planetesimals form ($\dot{M}=10^{-7}\,{\rm
  M_\odot\,yr^{-1}}$ and $\dot{M}=10^{-6}\,{\rm M_\odot\,yr^{-1}}$). The
  isolation mass by planetesimal accretion, with a feeding zone here assumed to
  be four Hill radii, is indicated by the dot-dashed lines. The protoplanets
  will grow to their isolation mass within approximately 1 Myr in all cases. For
  planetesimal formation in a dense protoplanetary disc, with
  $\dot{M}=10^{-6}\,{\rm M_\odot\,yr^{-1}}$, the isolation mass is similar to
  the mass of Mercury and no giant impacts are needed to explain its
  characteristic mass. The expected mass of planetesimals formed for each
  heating event of the protoplanetary disc is indicated by dash-triple-dotted
  lines.}
  \label{f:planetary_growth}
\end{figure}

We make here a simple calculation of the growth from planetesimal to protoplanet
via planetesimal accretion. The growth equation includes gravitational focusing,
\begin{equation}
  \dot{M} = \pi R^2 \varSigma_{\rm pla} \varOmega \left[ 1 + \left( \frac{v_{\rm
  esc}}{v_{\rm pla}} \right)^2 \right] \, .
\end{equation}
Here $\varSigma_{\rm pla}$ is the surface density of planetesimals, assumed to
be 0.01 times the surface density of gas when the planetesimals form.  We set
the planetesimal speed $v_{\rm pla} = e_{\rm pla} v_{\rm K}$ where $e_{\rm pla}$
is set by the balance between turbulent stirring and gas drag on planetesimals
(equation \ref{eq:edrag}).  We adopt a viscous protoplanetary disc model where
the gas accretion rate $\dot{M}_\star$ falls with time \citep[see][for
details]{Johansen+etal2019}. The resulting mass growth is shown in Figure
\ref{f:planetary_growth} for two different values for the planetesimal radius
(100 km and 10 km) and two different values for the gas accretion rate when the
planetesimals form ($\dot{M}_\star = 10^{-7}\,{\rm M_\odot\,yr^{-1}}$ and
$\dot{M}_\star = 10^{-6}\,{\rm M_\odot\,yr^{-1}}$). A planet with the mass of
Mercury can form by accretion of iron-rich planetesimals within 1 Myr under all
circumstances. For $\dot{M}_\star = 10^{-6}\,{\rm M_\odot\,yr^{-1}}$ at the
planetesimal formation stage, the planetesimal isolation mass is, intriguingly,
very similar to Mercury's mass. A lower surface density of planetesimals, as
exemplified here by the calculations using $\dot{M}_\star = 10^{-7}\,{\rm
M_\odot\,yr^{-1}}$, has isolation masses ten times lower than Mercury's mass and
would require an additional giant impact stage to read the full mass of Mercury.

The reservoir of iron-rich planetesimals that form at each heating event in the
protoplanetary disc can be assessed from the expression
\begin{equation}
  M_{\rm res} \sim \varSigma_{\rm pla} 2 \pi r \delta r \, ,
\end{equation}
where $\varSigma_{\rm pla}$ is the planetesimal column density and $\delta r$ is
the width of the zone where iron planetesimals form. We calculate
$\varSigma_{\rm pla}=0.01 \varSigma_{\rm g}$ from the gas surface density
$\varSigma_{\rm g}$ evaluated from the stellar mass accretion rate through
$\dot{M}_\star = 3 \pi \nu_{\rm t} \varSigma_{\rm g}$ (here $\nu = \alpha c_{\rm
s} H$ is the turbulent viscosity). Taking $\delta r = r = 0.4\, {\rm AU}$ we get
$M_{\rm res} \approx 0.36 \, M_{\rm E}$ for $\dot{M}_\star = 10^{-7} \, M_\odot
\, {\rm yr^{-1}}$ and $M_{\rm res} \approx 1.64 \, M_{\rm E}$ for $\dot{M}_\star
=10^{-6} \, M_\odot \, {\rm yr^{-1}}$. Thus, Mercury could form from the mass
reservoir of a single planetesimal formation event, while super-Mercuries in the
mass range 5--10 $M_{\rm E}$ would need to accumulate planetesimals from several
protoplanetary disc heating events at the earliest stages of protoplanetary disc
evolution where the stellar mass accretion rate is high. Very young stars are
estimated to undergo $\sim$$10$ FU-Orionis-like outbursts
\citep{HillenbrandFindeisen2015} and this yields a mass budget that allows the
formation of super-Mercuries by accretion of iron planetesimals.

\subsection{Latent heat}

We ignored the latent heat release during the vapour deposition. The ratio of
the heating by latent heat release to the cooling by radiation is
\begin{equation}
  \frac{\dot{E}_{\rm heat}}{\dot{E}_{\rm c}} = \frac{L \rho_{\rm v}
  v_\perp}{\sigma_{\rm SB} T^4} \, .
\end{equation}
Here $L$ is the latent heat, $\rho_{\rm v}$ is the vapour density and $v_\perp$
is the speed at which iron vapour impinges onto the surface of a growing
particles. For iron with $L=6.285$ MJ/kg, $T=1400$ K, $v_\perp = 400$ m/s and
$\rho_{\rm v}=10^{-8}$ kg/m$^3$ we get that latent heat contributes 0.1\%-1\% of
the heat relative to the cooling. Therefore latent heat is safely negligible.

\subsection{Calcium-aluminium rich planets}

\citet{Dorn+etal2019} suggest the presence of close-in super-Earths that are
enhanced in CAI material (ultrarefractory calcium-aluminium rich minerals,
represented in our model by the corundum mineral) and depleted in iron (e.g., 55
Cnc e, WASP-47 e).  These planets have characteristically lower densities (by
10-20\%) than the majority of super-Earths. It remains unclear how to obtain
reservoirs in the proto-planetary disc that allow to form CAI-rich planets of
several $M_{\rm E}$.  Previous formation models that employed N-body simulations
in combination with equilibrium chemistry models achieve the formation of up to
1 or 3 $M_{\rm E}$ \citep{Carter-Bond+etal2012, Thiabaud+etal2014}.

Our protoplanetary disc model (Figure \ref{f:sublimation_fronts}) has a
separation of approximately 0.3 AU between the iron sublimation front at an
ambient temperature of 1300 K and the corundum destruction front at a
temperature of approximately 1600 K. If planetesimals form in this region, then
they would be rich in elements present in refractory minerals (such as Al and
Ca). Several tens of Earth masses ultrarefractory material will flow through
this ``CAI region'' of the protoplanetary disc and onto the star during the
early phases of protoplanetary disc evolution when the stellar accretion rate is
high.

The low abundance of Al relative to Fe, Mg and Si (solar Al:Si=0.084)
gives very low condensational growth rates of corundum and these
minerals grow at most to a few 100 microns in size (or ${\rm St} \sim 10^{-4}$)
in our model, see Figure \ref{f:homogeneous_nucleation_rate}. Calcium-Aluminium
Rich Inclusions (CAI) found in meteorites are up to several cm in size
\citep{Krot+etal2004,Toppani+etal2006}. These large CAI nevertheless often
display characteristics of melting \citep{Charnoz+etal2015}, which may hide
their accretional history. More primitive CAI appear to be aggregates of nodules
of 10--50 microns in size \citep{Krot+etal2004}. Thus CAI aggregates large
enough to drive formation of ultrarefractory planetesimals might be achieved in
our model by coagulation of microscopic CAI-lets that formed by nucleation and
slow condensation. The efficiency of this growth within the MRI-unstable region
of the protoplanetary disc would nevertheless require a better understanding of
how such ultrarefractory minerals stick under high collisions speeds.

The pebble accretion efficiency (defined as the rate of pebble accretion onto a
planet divided by the flux of pebbles through the protoplanetary disc) increases
with both decreasing distance from the star as well as with decreasing Stokes
number \citep{LambrechtsJohansen2014}. Ultrarefractory planets would therefore
also experience significant pebble accretion rates. We nevertheless caution that
both planetesimal formation and pebble accretion in strongly turbulent regions
of the protoplanetary discs are poorly understood
\citep{Johansen+etal2007,Xu+etal2017,Gole+etal2020}. Our nucleation models can
be used in future work using $N$-body simulations to evaluate planetary growth
in the innermost regions of protoplanetary discs.

\subsection{Ultrarefractory metal seeds}

We have ignored the possibility that rare ultrarefractory metal grains (made
e.g.\ out of O) could act as substrates for heterogeneous nucleation of iron.
\cite{KozasaHasegawa1988} investigated this possibility and included that Os has
a very high surface tension and hence will nucleate after Fe, unless the cooling
time is very long. The abundance of Os is anyway very low and even if it
condenses first, it will likely form few and very small grains only. That would
still lead to formation of large metal pebbles.

\subsection{Validity of Classical Nucleation Theory}

\cite{DonnNuth1985} criticized the application of CNT to the low-density,
high-temperature environment in AGB stars. The CNT approach requires that the
collision time-scale is much shorter than the time-scale for changing the super
saturation. \cite{NozawaKozasa2013} compared CNT with non-steady models and
found that CNT is valid when the collision frequency of atoms on a cluster is
much higher than the rate of change of the supersaturation. These conditions are
safely fulfilled in the dense protoplanetary disc environment.

\cite{Kimura+etal2017} reported an extremely low sticking efficiency
($\alpha_{\rm s} \sim 10^{-5}-10^{-4}$) for homogeneous nucleation of iron from
the vapour phase in micro-gravity experiments. Various explanations for the
difference between these experiments and ground-based experiments showing much
higher sticking efficiencies are discussed in \cite{Kimura+etal2017}. As we
discussed in Section \ref{s:nucleation_model}, we already effectively assumed a
sticking efficiency of $\alpha_{\rm s} \sim 10^{-2}$ to match the
\cite{Kashchiev2006} model to the experimental nucleation data. We simply note
here that the sticking efficiency for the homogeneous nucleation of a wide range
of minerals under micro-gravity conditions would be needed to assess the outcome
of these experiments on the nucleation competition between the different
minerals.

\section{Conclusions}

We have explored in this paper the homogeneous nucleation of iron and other
refractory minerals following a protoplanetary disc heating event (such as an FU
Orionis outburst). Perhaps the most surprising outcome of our calculations is
that all these minerals nucleate at very high supersaturation levels. Metal and
rock have high surface tension and the minerals therefore do not benefit from
heterogeneous nucleation on existing surfaces. Homogeneous nucleation directly
from the gas phase requires extreme values of the supersaturation to overcome
the high surface tension of small clusters to grow to stable, nucleated
particles. Iron has the largest surface tension of the minerals considered here
(Al$_2$O$_3$, MgSiO$_3$, Fe, Fe+S and FeO) and nucleates at supersaturation
levels up to above $S = 10^5$.

The difficulty in nucleating new iron particles directly implies that iron
pebbles grow to very large sizes, up to a few cm at low cooling rates. This
large growth is due to the scarcity of nucleated nano-scale particles. The lucky
particles that do nucleate are therefore exposed to a large reservoir of iron
vapour to drive further growth to macroscopic pebbles. We have used enstatite
(MgSiO$_3$) to represent rock-forming minerals. Enstatite nucleates at lower
supersaturation levels than iron and therefore the pebbles grow to sizes that
are a factor few smaller than the iron pebbles. Combined with the intrinsic
material density difference, this gives a factor ten difference in Stokes number
between iron and enstatite pebbles.

We propose that this mechanism is at the heart of the curious composition of
Mercury. Planetesimals formed by the streaming instability preferably
incorporate pebbles with the largest Stokes number from the background size
distribution. These planetesimals therefore become iron-rich. We demonstrate
that accretion of iron-rich planetesimals leads to the formation of Mercury-mass
planets within a million years. The mass reservoir is large enough to form
iron-rich super-Earths if the protoplanetary disc undergoes several FU Orionis
outbursts during the earliest phases of protoplanetary disc evolution. The
iron-rich planetesimals will also contain a minor fraction of silicates.
Importantly, these silicates will be very reduced (poor in oxidized FeO) because
of the separation of Fe on one side and MgO/SiO$_2$ minerals on the other side
by the nucleation process. This may explain why Mercury's mantle has the lowest
FeO fraction of all rocky bodies in the inner Solar System
\citep{WarellBlewett2004}.

A growing number of exoplanets are known now to have a density similar to
Mercury's and covering a large range of masses up to $\sim$5--10 $M_{\rm E}$.
Their high density could arguably be a consequence of mantle stripping in giant
impacts, as has traditionally been used to explain Mercury's high core fraction.
Our model, in contrast, implies that the formation of this class of iron-rich
and highly reduced planets is a direct consequence of luminosity variations of
very young stars and a nucleation process that separates iron from silicates.

\begin{acknowledgements}

The authors would like to thank the anonymous referee for many valuable
comments that helped improve the manuscript. A.J.\ acknowledges funding from
the European Research Foundation (ERC Consolidator Grant 724687-PLANETESYS), the
Knut and Alice Wallenberg Foundation (Wallenberg Scholar Grant 2019.0442), the
Swedish Research Council (Project Grant 2018-04867), the Danish National
Research Foundation (DNRF Chair Grant DNRF159) and the G\"oran Gustafsson
Foundation. C.D.\ acknowledges support from the Swiss National Science
Foundation under grant PZ00P2\_174028.

\end{acknowledgements}

\end{document}